# First-principles study of the relaxor ferroelectricity of Ba(Zr,Ti)O$_3$[*]

Yang Li-Juan [a),b)], Wu Ling-Zhi [c)], and Dong Shuai [b)†]

[a)]*School of Information Science and Technology, Suqian College, Suqian 223800, China*
[b)]*Department of Physics, Southeast University, Nanjing 211189, China*
[c)]*School of Geography and Biological Information, Nanjing University of Posts and Telecommunications, Nanjing 210046, China*

Ba(Zr,Ti)O$_3$ is a lead-free relaxor ferroelectric. Using the first-principles method, the ferroelectric dipole moments for pure BaTiO$_3$ and Ba(Zr,Ti)O$_3$ supercells have been studied. All possible ion configurations of BaZr$_{0.5}$Ti$_{0.5}$O$_3$ and BaZr$_{0.25}$Ti$_{0.75}$O$_3$ are constructed in a 2×2×2 supercell. For the half-substituted case, divergence of ferroelectric properties has been found among these structures, which seriously depends on the arrangement of Ti and Zr ions. Thus our results provide a reasonable explanation to the relaxor behavior of Ba(Zr,Ti)O$_3$. In addition, a model based on the thermal statistics gives the averaged polarization for Ba(Zr,Ti)O$_3$, which depends on the temperature of synthesis. Our result is helpful to understand and tune the relaxor ferroelectricity of lead-free Ba(Zr,Ti)O$_3$.

[*] Project supported by the Natural Science Foundation of China (grant Nos. 51322206 and 11274060) and the Natural Science Foundation of Jiangsu Province (grant No.15KJB140009).
[†] Correspondence author. Email: sdong@seu.edu.cn

# 1. Introduction

The ABO$_3$-type perovskite oxides have been extensively studied, because of their novel physics and broad technical applications.[1] Among abundant functional perovskite compounds, ferroelectric materials, such as BaTiO$_3$ and PbTiO$_3$, have been known for a long time and are acquiring increasing importance in modern electronic industry where their unique properties are utilized in capacitors and piezoelectric transducers.[2] In addition, the partial Zr substituted PbTiO$_3$ (Pb(Zr,Ti)O$_3$) is widely used to produce the ferroelectric random access memory (FeRAM) and other devices[3] for the principal advantages of relative low crystallization temperature, compatibility with complementary metal oxide semiconductor (CMOS) and high remanent polarizations.[4] However, although PbTiO$_3$ and Pb(Zr,Ti)O$_3$ are very industrial valuable but they are environmentally unfriendly due to the toxicity of Pb. It remains a challenge problem to pursuit the Pb-free ferroelectrics to replace PbTiO$_3$/ Pb(Zr,Ti)O$_3$ in industry. In the past years, there are a few experiment studies on Ba(Zr,Ti)O$_3$ thin films, which can be considered as a clone of Pb(Zr,Ti)O$_3$. Although its ferroelectric performance is not so prominent comparing with Pb(Zr,Ti)O$_3$, it provides an alternative choice to environmental protection. Comparing with the intensively studied Pb(Zr,Ti)O$_3$, the investigations on Ba(Zr,Ti)O$_3$ are much less. Although there are a few experimental works on this material,[5-12] a systematic theoretical study on its physical mechanism of relaxor ferroelectricity of Ba(Zr,Ti)O$_3$ is absent, to our best knowledge. In fact, the physical issues involved in Ba(Zr,Ti)O$_3$ are not identical to Pb(Zr,Ti)O$_3$, since Pb$^{2+}$ itself contributes a significant portion of polarization. In this sense, a theoretical investigation on Ba(Zr,Ti)O$_3$ is physical meaningful and useful to its applications.

BaTiO$_3$ is a typical ferroelectrics perovskite. Its phase transition from paraelectric state to ferroelectric state has been experimentally studied using various techniques. With deceasing temperature, it has three sequent phase transitions, from cubic to tetragonal at 393 K, tetragonal to orthorhombic at 278 K, and orthorhombic to rhomohedral at 183 K.[13,14] The paraelectric phase is cubic (space group *Pm3m*) with the experimental lattice constant of 4.0045 Å.[14,15] Among all the ferroelectric states, the tetragonal ferroelectric state is most interesting and has been widely studied previously.[16] The tetragonal ferroelectric phase at room temperature belongs to the space group *P4mm*. The experimental measured lattice constants are: *a*=3.9945 Å and *c*=4.0335 Å. The fractional atomic positions are: Ba at (0, 0, 0), Ti at (0.5, 0.5, 0.514), O$_1$ at (0.5, 0.5, -0.025), O$_2$ at (0.0, 0.5, 0.488) and O$_3$ at (0.5, 0.0, 0.488).[14,17] And its spontaneous polarization ($P_S$) is 26 μC/cm$^2$ along the tetragonal *c*-axis.[18,19] The partial substitutions of Zr occupies the B-sites (Ti's site) randomly.

In this work, the ferroelectricity of Ba(Zr,Ti)O$_3$ will be studied with the supercell model by the first-principles theory. Our density functional theory (DFT) calculations find that in general the dipole moment decreases with Zr's substitution. The ferroelectric properties diverge among different structures, which seriously depend on the arrangement of Ti and Zr ions. In addition, the model based on thermal statistics revealed the synthesis-temperature dependent polarization of Ba(Zr,Ti)O$_3$. Our theoretical results can explain some experimental observations like relaxor ferroelectricity and help to improve its ferroelectric performance.

## 2. Method

Our DFT calculations were performed using Vienna *ab initio* Simulation Package (VASP)[20, 21] with the PBEsol (Perdew-Burke-Ernzerhof-revised)[22] parametrization of the generalized gradient approximation plus *U* (GGA+*U*).[23-25] The choice of PBEsol is crucial to give accurate description of the lattice structure of titanates, while the traditional PBE is often failed for ferroelectric titanates. All calculations, including the lattice relaxation and static calculations, have been done with the Hubbard $U_{eff}$ (=*U*-*J*) applied on the *d*-orbitals of Ti ion using the Dudarev implementation.[26] The plane-wave cutoff of 550 eV is adopted. A 9×9×9/5×5×5 Monkhorst-Pack *k*-point sample centered at the Γ point is used in combination with the tetrahedron method for the pure BaTiO$_3$ and the 2×2×2 supercell respectively. The inner atomic positions as well as the lattice constants are fully optimized as the Hellman-Feynman forces are converged to be less than 5 meV/Å. The electrical polarization is calculated using the standard Berry-phase approach.[27]

## 3. Results and discussion

Before the calculation of Ba(Zr,Ti)O$_3$, the physical properties of BaTiO$_3$ are checked first. The tetragonal structure of ferroelectric state is fully optimized with various $U_{eff}$ from 0 eV to 4 eV stepped by 1 eV. Then the electric dipole moment and band gap are calculated using the fully relaxed structures. As shown in Fig. 1, the dipole moment varies from 30.27 μC/cm$^2$ to 1.21 μC/cm$^2$ and the band gap increases with increasing $U_{eff}$. When $U_{eff}$ =1 eV, the fully relaxed lattice gives: $a$=3.988 Å, $c$=4.044 Å (Fig. 1(c)) and the ferroelectric dipole moment is just 25.15 μC/cm$^2$ (Fig. 1(a)) which are fairy close to the experimental data, implying a proper $U_{eff}$ for BaTiO$_3$. Although the band gap of ~2.0 eV (Fig. 1(b)) is little narrower than the experiment valence (2.8 to 3 eV[28]), it remains acceptable considering the well-known underestimation of band gap in DFT calculations. Also, we checked the suitable $U_{eff}$ for Zr ion. In the DFT calculation, the optimized lattice constant for cubic BaZrO$_3$ increases with $U_{eff}$. When $U_{eff}$ =0 eV, the lattice constant gives 4.215 Å, which is closest to the experiment value (4.19 Å).[29] This is also reasonable considering Zr's 4$d$ orbitals whose Hubbard $U$ is much weaker than the 3d orbitals. Thus, in the following calculations, $U_{eff}$=1 eV and 0 eV will be adopted by default for Ti's and Zr's $d$ orbitals, respectively.

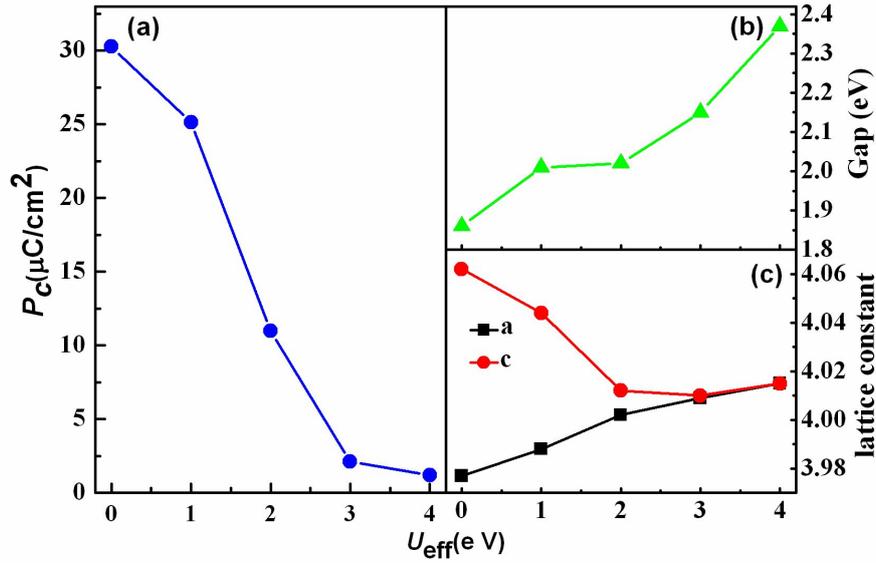

**Fig. 1.** (color online) Physical properties of BaTiO$_3$ in our DFT calculation as a function of $U_{eff}$. (a) The dipole moment. (b) The band gap. (c) The lattice constants.

Subsequently, the energies and dipoles for paraelectric and ferroelectric BaZr$_{0.5}$Ti$_{0.5}$O$_3$ are calculated using a 2×2×2 supercell which includes eight chemical units. Since in real materials Zr and Ti may be randomly distributed, here all possible Zr-Ti configurations are tested in the supercell. The combination number is $C_8^4=70$. Considering the space symmetry, there are only fifteen independent configurations for the ferroelectric state, which are further reduced to five independent configurations for the paraelectric state, which are summarized in Table 1. The degeneracy of these configurations are also presented.

**Table 1**. (color online) The DFT results of all models for BaZr$_{0.5}$Ti$_{0.5}$O$_3$ supercell. In the 2x2x2 supercell, the red balls are titanium ions, and the yellow

balls are zirconium ions. The barium and oxygen ions are not shown here. The energy difference $\Delta E$ in unite of meV between various paraelectric models: $E(x)-E(V)$ and ferroelectric ones: $E(x)-E(O)$ for the supercell. The dipole moments ($P_c$) along the c-axis for fifteen independent ferroelectric states are also shown in unit of $\mu C/cm^2$.

| paraelectric | | | ferroelectric | | | |
|---|---|---|---|---|---|---|
| Model | | $\Delta E$ | Model | Degeneracy | $\Delta E$ | $P_c$ |
| I | 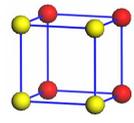 | 846 | A 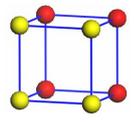 | 4 | 633 | 21.52 |
| | | | B 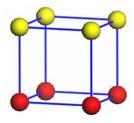 | 1 | 846 | 1.85 |
| | | | C 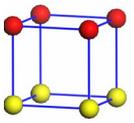 | 1 | 847 | 1.96 |
| II | 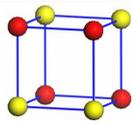 | 546 | D 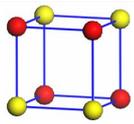 | 4 | 545 | 0.60 |
| | | | E 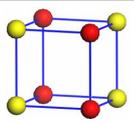 | 2 | 212 | 23.73 |
| III | 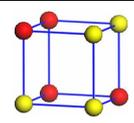 | 715 | F 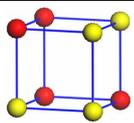 | 8 | 652 | 12.84 |
| | | | G 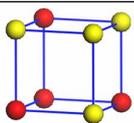 | 4 | 589 | 12.85 |
| | | | H 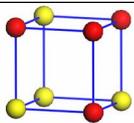 | 4 | 589 | 12.87 |

| | | | | | | | |
|---|---|---|---|---|---|---|---|
| IV | 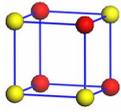 | 498 | I | 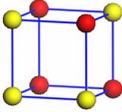 | 8 | 403 | 13.42 |
| | | | J | 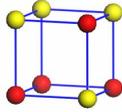 | 4 | 500 | 2.14 |
| | | | K | 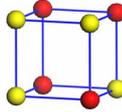 | 8 | 403 | 12.92 |
| | | | L | 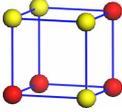 | 8 | 625 | 14.60 |
| | | | M | 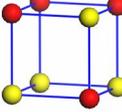 | 4 | 500 | 2.21 |
| | | | N | 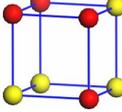 | 8 | 625 | 14.71 |
| V | 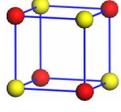 | 0 | O | 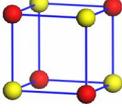 | 2 | 0 | 0.21 |

For each independent configuration, both the internal atoms' positions and the lattice framework are relaxed to get the optimal structure. Then the energy and dipole moment for the relaxed structure are calculated, as shown in Table 1 too. According to Table 1, among all configurations, the one named V/O has the lowest energy despite the paraelectric/ferroelectric state, in which Zr and Ti are alternatively ordered along all three axes. Its relaxed lattice gives: $a=b=8.204$ Å, $c=8.205$ Å for the ferroelectric state, whose tetragonality (namely the $c/a$ ratio) is very weak. As expected, its dipole moment along the $c$-axis is only 0.21 µC/cm², which is a very tiny value. As shown in Table 1, the calculated dipole moments are very divergent among different configurations, which seriously depend on the Zr-Ti arrangement. The maximum value is 23.73 µC/cm$^2$ for the configuration E, which is very close to the BaTiO$_3$ itself. Thus, our result suggest that for real Ba(Zr,Ti)O$_3$, the quenching disorder of Zr and Ti will make the system to be quite non-uniform regarding its local ferroelectric properties, rendering the relaxor ferroelectric behavior. [30, 31]

In order to obtain the effective polarization of real Ba(Zr,Ti)O$_3$, the thermal average based on the Boltzman equation (($p$~D·exp(-$E$/$k_B T$) where $p$ is the probability of each independent configuration, with the degeneracy $D$, $E$ is the energy, $k_B$ is the Boltzman constant, and $T$ is temperature) is performed to simulate the distribution of Zr-Ti configuration during the high-temperature synthesis. After the thermal equilibrium at the high temperature reaction, the crystal forms and then is cooled down to room temperature. Then the distribution of Zr-Ti configuration is modeled to be determined by the synthesis temperature. At such a high synthesis temperature, Ba(Zr,Ti)O$_3$ is in its paraelectric state since the ferroelectric Curie temperature of BaTiO$_3$ is only 393 K,[14] much lower than typical synthesis temperatures which can be higher than 1000 K.

Then the probabilities for five paraelectric states are calculated as a function of the synthesis temperature. As shown in Fig. 2(a), the lowest energy state V is dominant when the synthesis temperature is relative low. With increasing temperature, the portion of the second lowest energy state IV increases significantly and even overcomes the V state finally, since the degeneracy of IV is larger than that of V. Then the contribution of polarization from each configuration is calculated, considering the probability of corresponding ferroelectric state and its dipole moment. As shown in Fig. 2(b), it is clear that in the whole temperature range the primary source state is from the configuration IV, while other states are all negligible to ferroelectric polarization. As shown in Fig. 2(c), the total polarization increases almost linearly from 0.93 μC/cm$^2$ to 6.74 μC/cm$^2$ with increasing synthesis temperature from 1000 K to 2000 K. Our result suggests that a higher synthesis temperature can promote the effective polarization of Ba(Zr,Ti)O$_3$.

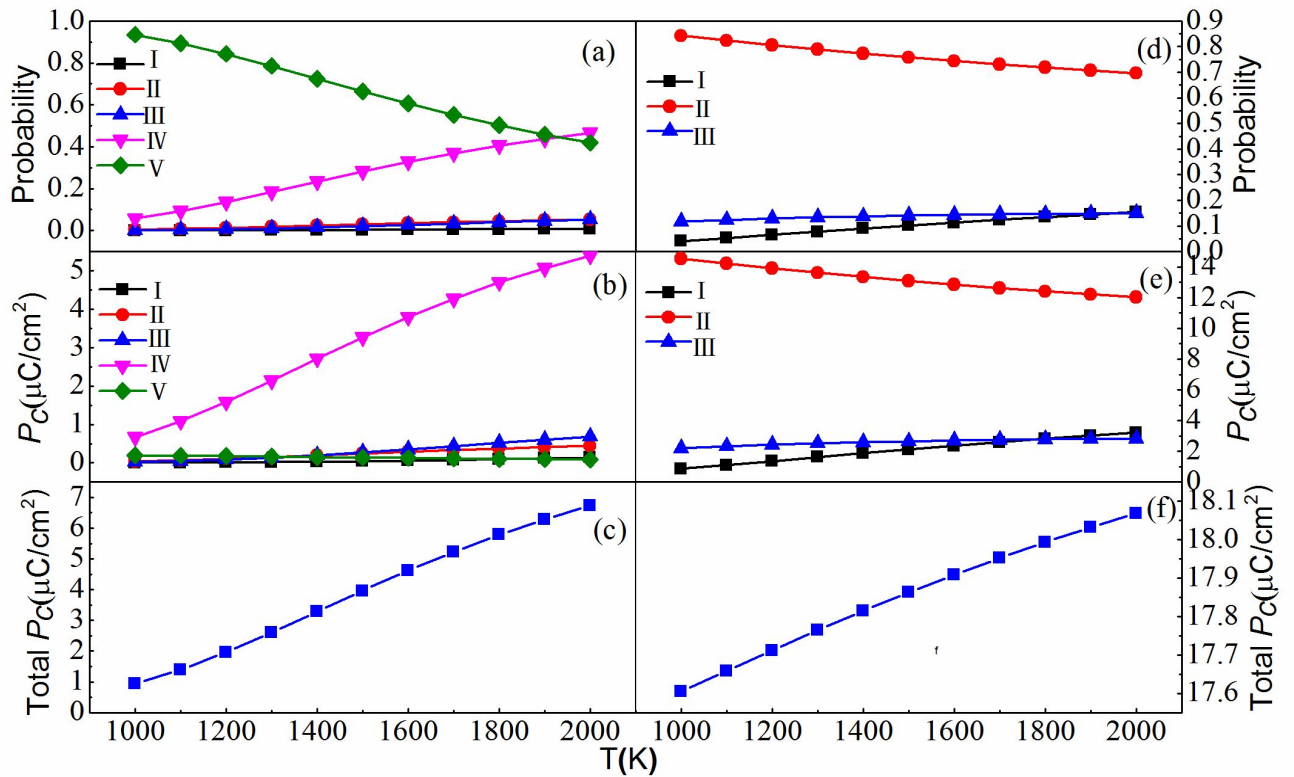

**Fig. 2.** (color online) (a-c) BaZr$_{0.5}$Ti$_{0.5}$O$_3$ supercell and (d-f) BaZr$_{0.25}$Ti$_{0.75}$O$_3$ supercell. (a) and (d) Probabilities for various paraelectric states as a function of synthesis temperature. (b) and (e) The dipole moments contribution of various

ferroelectric states evolved from corresponding paraelectric states as a function of synthesis temperature (c) and (f) Total dipole moments for the composites as a function of synthesis temperature.

According to above results on BaZr$_{0.5}$Ti$_{0.5}$O$_3$, the system becomes relaxor ferroelectric upon the partial substitution of Zr. However the effective polarization is greatly suppressed. Therefore, to keep a balance between the relaxor and polarization, less substitution may be a better choice. In the following, the concentration of Zr is reduced to 25%, giving BaZr$_{0.25}$Ti$_{0.75}$O$_3$. Still using the 2×2×2 supercell, despite the total $C_8^2=28$ combinations, there are seven independent configurations for the ferroelectric state and only three independent configurations for the paraelectric state, as summarized in Table 2. By performing the same processes (relaxation and static calculation), the energies and dipole moments for relaxed structures are obtained, which are also shown in Table 2. According to Table 2, the configuration II has the lowest energy in the paraelectric state and its corresponding ferroelectric branch. Different from above half-substituted case, the configuration B processes the maximum dipole moment (28.79 uC/cm$^2$) along the $c$-axis. Even the minimum dipole moment reaches 16.65 μC/cm$^2$ as in the configuration E. Then the probability of three paraelectric configuration, their corresponding ferroelectric contribution, and the total polarization are calculated as a function of synthesis temperature. According to Fig. 2(d-f), in the whole temperature range from 1000 K to 2000 K, the configuration II is the dominant one and contributes most polarization. The total ferroelectric polarization only slightly changes from 17.64 μC/cm² to 18.06 μC/cm² with increase synthesis temperature. In this sense, the quarter-substituted Ba(Zr,Ti)O₃ is more like a normal ferroelectric while its relaxor behavior is not prominent.

**Table 2.** (color online) The DFT results of all models for BaZr$_{0.25}$Ti$_{0.75}$O$_3$ supercell.

| paraelectric | | | ferroelectric | | | |
|---|---|---|---|---|---|---|
| Model | | ΔE | Model | | Degeneracy | ΔE | P$_c$ |
| I | 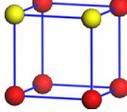 | 260 | A | 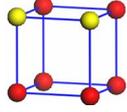 | 4 | 299 | 16.97 |
| | | | B | 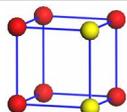 | 4 | 152 | 28.79 |
| | | | C | 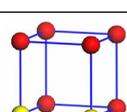 | 4 | 299 | 16.99 |
| II | 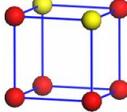 | 0 | D | 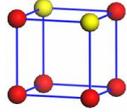 | 2 | 0.04 | 18.49 |
| | | | E | 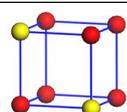 | 8 | 40 | 16.65 |
| | | | F | 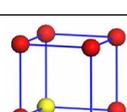 | 2 | 0 | 18.52 |
| III | 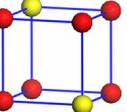 | 76 | G | 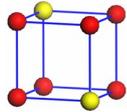 | 4 | 76 | 18.88 |

In summary, the relaxor ferroelectric Ba(Zr,Ti)O₃ supercells have been studied based on the density functional theory and thermal statistics. For the half-substituted case, divergence of ferroelectric dipole moments has been found among different Zr-Ti configurations. Thus, the well known relaxor behavior of Ba(Zr,Ti)O$_3$ can be attributed to the quenching disorder which gives rise to the non-uniform ferroelectric properties. In addition, our calculation indicates that the effective total polarization increases with the synthesis temperature, which may be helpful to improve the ferroelectric performance of real material. In contrast, the quarter-substituted Ba(Zr,Ti)O$_3$ shows different behavior, which is more like a normal ferroelectric instead of a relaxor.